\documentclass[11pt,dvips]{article}
\usepackage{amssymb}
\usepackage{graphicx}
\usepackage{amsmath}  

\newcommand{\ha}{\mbox{\small$\frac{1}{2}$}}
\newcommand{\qu}{\mbox{\small$\frac{1}{4}$}}
\newcommand{\im}{\,{\rm i}\,}
\newcommand{\lab}[1]{\label{#1}}
\newcommand{\re}[1]{(\ref{#1})}
\newcommand{\nn}{\nonumber}

\newcommand{\B}[1]{\mbox{\boldmath$#1$}}
\newcommand{\sB}[1]{\mbox{\scriptsize\boldmath$#1$}\vphantom{#1}}
\newcommand{\Blambda}{\mbox{\boldmath$\lambda$}}
\newcommand{\D}[2]{{\rm d}^{#1}{#2}\,}
\newcommand{\inta}{\hspace*{-.1cm}\int \hspace*{-.15cm}}

\def\ds{\displaystyle}

\def\di{{\partial}}
\def\ie{{\it i.e.~}}

\def\be{\begin{equation}}
\def\ee{\end{equation}}
\def\bea{\begin{eqnarray}}
\def\eea{\end{eqnarray}}
\def\Ft{{\cal F}}
\def\L1{{\cL_{(1)}}}

\def\cL{{\cal L}}
\def\cA{{\cal S}}
\def\bx{{\bf x}}

\def\Aslash{A\kern-0.47em/}
\begin{document}
\title{Analysis of inter-quark
interactions in classical chromodynamics}

\author{J. W. Darewych$^1$ and A. Duviryak$^2$\cr
{\it $^1$Department of Physics and Astronomy,}\cr
{\it York University, Toronto, ON M3J 1P3, Canada}\cr
{\it $^2$Department for Computer Simulations of Many-Particle Systems,}\cr
{\it Institute for
Condensed Matter Physics of NAS of Ukraine, Lviv, UA-79011, Ukraine}
}
\date{\small
15 XII 2011}
\maketitle
\begin{abstract}
The QCD gluon equation of motion is solved approximately by means of
the Green function. This solution is used to reformulate the
Lagrangian of QCD such that the gluon propagator appears directly in
the interaction terms of the Lagrangian. The nature of the
interactions is discussed. Their coordinate-space form is presented
and analyzed in the static, non-relativistic case.
\end{abstract}


\section{QCD Lagrangian and Equations of Motion}
\renewcommand{\theequation}{1-\arabic{equation}}
\setcounter{equation}{0}

The QCD Lagrangian density is \cite{PLB} ($\hbar=c=1$)
%
\begin{equation}
\label{1-1} {\cal L}_{QCD}=-{1\over
4}F^{(a)\,\mu\,\nu}\,F_{\mu\,\nu}^{(a)} + \sum_q \,{\bar \psi}^i_q
\, \big( \im  \gamma^{\mu}\,(D_{\mu})_{i\,j} - m_q \, \delta_{i
j}\big)\,\psi^j_q
\end{equation}
where
%
\begin{equation} \label{1-2}
F^{(a)}_{\mu\,\nu}=\partial_{\mu}\,A^{(a)}_{\nu}-\partial_{\nu}\,A^{(a)}_{\mu
}-g_s\,f_{a\,b\,c}\,A^{(b)}_{\mu}\,A^{(c)}_{\nu},
\end{equation}
%
and
%
\begin{equation} \label{1-3}
(D_{\mu})_{i\,j}=\delta_{i\,j}\,\partial_{\mu}+\im\,g_s\,t^{(a)}_{i\,j}\,A^{(a)}_{\mu}
~~~~~{\rm or}~~~~ {\bf D}_\mu = {\bf I}_{3 \times 3}\,
\partial_{\mu} + \im \, g_s \,  A^{(a)}_\mu \, {\bf t}^{(a)}.
\end{equation}
The QCD coupling constant is $g_s$, $f_{a\,b\,c}$ are
the structure constants, and
%
\begin{equation} \label{1-4}
t^{(a)}_{i\,j}= \frac{1}{2} \, \lambda^{(a)}_{i\,j},
\end{equation}
where the $\lambda$s are the Gell-Mann matrices. As usual, repeated
indices are summed over, with the colour indices $i,j=1,2,3$, the
flavour indices $q=d,u,s,c,b,t$,
and $a,b,c=1,...,8$ for the gluon fields.
Gluon indices will be usually indicated by round brackets to avoid
confusion with vector indices; eg. $A^{(a)}_\mu$ or $A^{(a)\,\mu}.$
%
\par      
It is convenient to write
%
\be \label{1-5} F^{(a)}_{\mu\,\nu}=\Ft^{(a)}_{\mu\,\nu} + G_{\mu \,
\nu}^{(a)} \ee
where
%
\be \label{1-6} \Ft^{(a)}_{\mu\,\nu} =
\partial_{\mu}\,A^{(a)}_{\nu}-\partial_{\nu}\,A^{(a)}_{\mu} \ee
is the free gluon field tensor and
%
\be \label{1-7} G_{\mu \,
\nu}^{(a)} = -g_s\;f_{a\,b\,c}\,A^{(b)}_{\mu}\,A^{(c)}_{\nu} \ee
is the non-Abelian part.
\par
Upon substituting \re{1-3} into \re{1-1}, the Lagrangian density
(\ref{1-1}) can be written as
%
\begin{equation} \label{1-8}
{\cal L}_{QCD}= {\cal L}_{\psi} + {\cal L}_{\Ft} +  {\cal L}_{I_1} +
{\cal L}_{I_3} + {\cal L}_{I_4}
\end{equation}
where
%
\begin{eqnarray}
\label{1-9}
&&{\cal L}_\psi = \sum_q {\bar \psi}^j_q \, ( \im \gamma^\mu \, \di_\mu - m_q) \, \psi^j_q \\
\label{1-10}
&&{\cal L}_{\Ft}=-{1\over 4}\, \Ft^{(a)\;\mu\,\nu}\;\Ft_{\mu\,\nu}^{(a)},\\
\label{1-11}
&&{\cal L}_{I_1}= - j_\psi^{(a)\, \mu} \, A^{(a)}_\mu ~~~~
{\rm where} ~~ j_\psi^{(a)\, \mu} =
g_s\,\sum_q {\bar \psi}^i_q \, t^{(a)}_{i\,j}\,\gamma^\mu \, \psi^j_q,\\
\label{1-12}
&&{\cal L}_{I_3} = -\frac{1}{2}\Ft^{(a)\;\mu\,\nu}\, G^{(a)}_{\mu\,\nu}
 = \frac{1}{2}\, g_s \, f_{a\,b\,c} \left(\di_\mu A^{(a)}_\nu - \di_\nu A^{(a)}_\mu \right)
 A_{(b)}^\mu \, A_{(c)}^\nu  = g_s f_{a\,b\,c} A_{(a)}^\mu \, A_{(b)}^\nu \di_\mu A^{(c)}_\nu ,~~~~~ \\
\label{1-13}
&&{\cal L}_{I_4} = -\frac{1}{4}\, G^{(a)\,\mu\,\nu}\, G_{\mu\,\nu}^{(a)}
= -\frac{1}{4} g_s^2\,  f_{a\,b\,c} \, f_{a\,d\,e}\, A^{(b)}_\mu \, A^{(c)}_\nu
\, A_{(d)}^\mu \, A_{(e)}^\nu.
\end{eqnarray}
\par
For the quark equations of motion, we need
%
\begin{equation} \label{1-14}
{\partial{\cal L}\over {\partial \bar\psi^{i}_q}}- {\di \over {\di
x^{\mu}}} \; \left( {\di {\cal L}\over {\di \bar\psi^i_{q, \mu}}}
\right)=0
\end{equation}
and using (\ref{1-9}) and (\ref{1-11}) we acquire the colour Dirac equations:
%
\begin{equation}   \label{1-15}
\left(\im\gamma^{\mu}\,\partial_{\mu}-m_q\right)\psi^i_q-g_s\,t^{(a)}_{i\,j}\,{\Aslash}^{(a)}\psi^j_q=0.
\end{equation}
Equations \re{1-15} can be written as
%
\begin{equation}  \label{1-16}
\im \gamma^{\mu} \, \left(D_{\mu}\right)_{ij} \psi^j_q - m_q
\psi^i_q = 0,
\end{equation}
or, in matrix notation,
%
\begin{equation}  \label{1-17}
\im\gamma^{\mu} \, {\bf D}_{\mu} \Psi_q - m_q \Psi_q = 0,
\end{equation}
where ${\bf D}_{\mu} = \left[\left(D_{\mu}\right)_{ij}\right]$ is
defined in equation (\ref{1-3}) and $\Psi_q^T = \left[\psi_q^1,
\psi_q^2, \psi_q^3\right]$ \; ($T$ stands for transpose).
\par
The equation of motion for the gluon fields is
%
\begin{equation}  \label{1-18}
{\bf D}^{\mu} F_{\mu \nu}^{(a)} = g_s \, \sum_q {\bar \Psi}_q {\bf t}^{(a)} \gamma_\nu \Psi_q,
\end{equation}
%
where ${\bf t}^{(a)} = \left[ t^{(a)}_{ij}\right]$ (cf. ref. \cite{DGH}). In component form, eq. \re{1-18} is
(cf. ref. \cite{PascTarr})
%
%
%
\begin{equation} \label{1-19}
\di^\mu \Ft^{(a)}_{\mu\,\nu}  = \partial _{\mu }\partial ^{\mu }A_\nu^{(a)}(x)-\partial_{\nu }\partial^{\mu
}A_\mu^{(a)}(x)=j_\nu^{(a)}(x)
\end{equation}
where
%
\begin{equation} \label{1-20}
j^{\nu \,(a)}(x)= j^{\nu \,(a)}_\psi (x) + j^{\nu \,(a)}_g (x) + j^{\nu \,(a)}_{g^2}(x) ,
\end{equation}
and
%
\begin{equation} \label{1-21}
j^{\nu \,(a)}_\psi = \sum_q \, g_s \, \bar{\psi }^i_q \, \gamma ^{\nu } \, t^{(a)}_{ij} \, \psi^j_q ,
\end{equation}
\begin{equation} \label{1-22}
j^{\nu \,(a)}_g = g_s \, f_{abc}\, \big(A^\nu_{(c)} \, \di_\mu A^\mu_{(b)} + 2\, A^\mu_{(b)} \, \di_\mu A^\nu_{(c)}
+ A^\mu_{(c)} \,\di^\nu A_\mu^{(b)}\big) ,
\end{equation}
\begin{equation} \label{1-23}
j^{\nu \,(a)}_{g^2} = g_s^2 \, f_{abc}\, f_{bde}\, A_\mu^{(c)} A^\mu_{(d)} \,  A^\nu_{(e)} .
\end{equation}
%
%

\section{Reformulation}
\renewcommand{\theequation}{2-\arabic{equation}}
\setcounter{equation}{0}

For the study of inter-quark interactions
(and subsequently for the study of the properties of mesons and baryons)
it is convenient to use a (formal) solution of the gluon equations of motion \re{1-19}
to reformulate the Lagrangian, and thus the  action, 
of QCD, so that the gluon
propagator appears directly in the interaction terms. 
Such reformulation has been shown to be useful for the study of inter-particle forces
in scalar theory with a nonlinear mediating field \cite{DD1,DD2}.
\par
The formal solution of \re{1-19} involves the use of the symmetric Green function of that equation,
and this requires a choice of gauge. We shall use the Lorentz gauge,
$\partial _{\mu}A^{\mu \,(a) }(x) = 0$, whereupon the ``glue" equation \re{1-19} can be rewritten
as an integral equation,
%
\begin{equation} \label{2.1}
A^{\mu \,(a)}(x)=
\int \D{4}{x^{\prime}} \,
D(x-x^{\prime}) \, j^{\mu \, (a)}(x^{\prime }),
\end{equation}
where $D(x-x^{\prime })$ is a Green function defined by
%
\begin{equation} \label{2.2}
\partial _{\nu }\partial ^{\nu }D(x-x^{\prime })=
\delta^{4}(x-x^{\prime }) .
\end{equation}
The Green function can be written as
%
\begin{equation} \label{2.3}
D(x-x^{\prime }) = \int \frac{\D{4}{k}}{(2\pi)^4 } \, D(k)\,
e^{-i k \cdot (x-x^{\prime })},
\end{equation}
where $D(k) = {\cal P} / k^2$ is the momentum-space representation of the Green function.
(Note: We use the notation that $x, \, k$ stand for 4-vectors, i.e. $x = (t, {\B x})$ and
$k = (k^0, {\B k})$, etc.)
\par
We have not included the free-gluon solution of the homogeneous equation (\ref{1-19})
in (\ref{2.1}), since free gluons do not arise and so free-gluon solutions
will play no role in the present considerations.
\par
The expression (\ref{2.1}) is only a formal solution of (\ref{1-19}) (in the Lorentz gauge),
because the components $j^{\nu \,(a)}_g  + j^{\nu \,(a)}_{g^2}$ of the current
$j^{\nu \,(a)}$ depend on $A^{\mu \, (a)}$. Unfortunately, it is not possible to obtain an explicit, closed-form
solution of the non-linear equation \re{2.1} (\ie of \re{1-19}) for $A^{\mu (a)}$ in terms of the quark
fields $\psi^i_q$ (at least we do not know how to do so). Thus, one must resort to approximation methods.
\par
Equation  (\ref{2.1}) can be solved as an iterative
series for $A^{\mu \, (a)}$ (cf. ref. \cite{ShpDar02}). The
first order term in this sequence is just the expression (\ref{2.1}), but with $j^{\mu \, (a)}$
replaced by $j^{\mu \, (a)}_\psi$ only, that is
%
\be \label{2-4}
A^{\mu \,(a)}_{(1)}(x)= \int \D{4}{x^{\prime}} \, D(x-x^{\prime}) \, j^{\mu \, (a)}_\psi(x^{\prime })
= \int \D{4}{x^{\prime}} \, D(x-x^{\prime}) \,
\sum_q \, g_s \, \bar{\psi }^i_q (x') \, \gamma ^{\nu }  \, t^{(a)}_{ij} \, \psi^j_q (x') .
\ee
The index $(1)$ in \re{2-4}, as in all that follows, indicates that it is a first-order iterative
expression. We note that the gluon fields are expressed here explicitly in terms
of the quark fields $\psi^i_q$ and the propagator $D$ only.
\par
Correspondingly, to first order, this modifies the Lagrangian density \re{1-8}, to
%
\be \label{2-6}
\L1 = \cL_\psi  - \frac{1}{2} j^{\mu\, (a)}_\psi \, A^{(a)}_{(1) \, \mu}
+ \cL_{I_3}\left(A^{\nu \,(b)}_{(1)}\right)  + \cL_{I_4}\left(A^{\nu \,(b)}_{(1)}\right),
\ee
where, in \re{2-6}, $A^{\nu \,(b)}_{(1)}$ is as given in \re{2-4}.

In obtaining $\L1$, eq. \re{2-6}, we have used the fact that, in the Lorentz gauge,
%
\be \label{2-7}
\cL_\Ft = - \frac{1}{2} \di_\mu A^{(a)}_\nu \, \di^\mu A^{\nu \, (a)} \simeq
\frac{1}{2} A^{(a)}_\nu \, \di_\mu \di^\mu A^{\nu \, (a)},
\ee
where $\simeq$ means equality modulo surface terms.  Then, because
$\ds \di_\mu \di^\mu A^{\nu \, (a)} = j^{\nu \, (a)}_\psi$
in first order,
this means that  $\ds \cL_\Ft  = \frac{1}{2} A^{(a)}_\nu \, j^{\nu \, (a)}_\psi$ (in first order).
Thus, in light of \re{2-4}, the reformulated Lagrangian
density $\L1$ (eq. \re{2-6}), and so the corresponding Hamiltonian and the action,
is a functional of the gluon Green function $D(x-x')$ and the quark fields $\psi^i_q$ only. We shall refer
to the theory based the Lagrangian density $\L1$ as the ``reduced model".
\par
At this point we might mention that in the QED (or $SU(1)$) case, the non-Abelian terms $\cL_{I_3}$ and
$\cL_{I_4}$ do not arise, and $\L1$ corresponds to the reduced QED Lagrangian. This reduced QED Lagrangian
was used previously (in the quantized Hamiltonian formalism) to derive relativistic few-fermion equations
and, from them, the relativistic energy spectra for all bound states of positronium \cite{TD1}, muonium \cite{TD2}
and negative positronium and muonium ions \cite{BD}. The results are exact to $O(\alpha^4$), including, for positronium,
the field-theoretic virtual annihilation correction.
%
%
\section{Analysis of the inter-quark interactions in the reduced model including the static limit}
\renewcommand{\theequation}{3-\arabic{equation}}   
\setcounter{equation}{0}

We denote the action by $\cA = \ds \int \D{4}{x} \, \cL (x)\,$ and shall consider,
in turn,  each of the  interaction terms of the action in the reduced model.
Thus, the term corresponding to $\cL_{I_1}$, eq. \re{1-11}, is
%
\be \label{3.1}
\cA ^{(1)}_{I_1} = \int d^4x \, \left(- \frac{1}{2} \, j^{\mu\, (a)}_\psi (x) \, A^{(a)}_{(1)\, \mu}(x)\right)
 =  - \frac{1}{2} \, \int \D{4}{x} \, \D{4}{x'}\, j^{\mu\, (a)}_\psi (x)\, D(x-x')\, j^{(a)}_{\psi \; \mu} (x),
\ee
where the quark-field current $j^{\mu\, (a)}_\psi (x)$ is given in eq. \re{1-11} \big(also in
\re{1-21}\big). Since  $j^{\mu\, (a)}_\psi  \propto g_s$, we see that $\cA ^{(1)}_{I_1}$ corresponds to an energy
contribution of $O(g_s^2)$. We also see that this interaction term describes the quark currents interacting
 via the gluon Green function $D$, which, in the quantised theory, would correspond to a one-gluon exchange
inter-quark interaction.
\par
To understand the physical content of the interactions corresponding to $\cA ^{(1)}_{I_k} (k=1,2,3)$,
it is useful to consider the
case of static sources in the non-relativistic limit, as was done for the scalar model
\cite{DD1,DD2}.
For the static case, $j_\psi^{\mu \, (a)} (x) = j_\psi^{\mu \, (a)} (\B x)$,
hence the static version of $\cA ^{(1)}_{I_1}$, eq. (\ref{3.1}), is
%
\be \label{3.2}
 \cA^{(1)}_{I_1} = - \frac{1}{2} \int \D{}t \, \D{}{t'} \, \D{3}x \, \D{3}x' \,
j^{\mu\, (a)}_\psi (\B x) \, D(x-x') \, j^{(a)}_{\psi \; \mu} (\B x') .
\ee
Since
%
\be \label{3.3}
D(x-x') =  \frac{1}{\pi} \delta\left((x-x')^2\right) = \frac{1}{8\pi} \frac{1}{|\B x-\B x'|}
\big[\delta(t-t'-|\B x-\B x'|) + \delta(t-t'+|\B x-\B x'|)\big]
\ee
it follows that
%
\be \label{3.4}
\int \D{}{t'}\, D(x - x') = \frac{1}{8\pi} \frac{1}{|\B x-\B x'|}
 \int \D{}{t'} \, \big[\delta(t-t'-|\B x-\B x'|) + \delta(t-t'+|\B x-\B x'|)\big]
= \frac{1}{4\pi} \frac{1}{|\B x-\B x'|}.
\ee
Thus, in the static case, $\cA ^{(1)}_{I_1}$, eq.  \re{3.1}, can be written as
%
\be \label{3.5}
 \cA ^{(1)}_{I_1} =  \int \D{}t \, L^{(1)}_{I_1}  = - \int \D{}t \, H^{(1)}_{I_1}
 ~~~{\rm where}~~ H^{(1)}_{I_1} =  \frac{1}{2} \int  \D{3}{x} \,  \D{3}{x'}
\, j^{\mu\, (a)}_\psi (\B x) \, \frac{1}{4\pi} \frac{1}{|\B x-\B x'|} \, j^{(a)}_{\psi \; \mu} (\B x')
\ee
It is clear from \re{3.5} that $H^{(1)}_{I_1}$ corresponds to the two-point potential energy function
%
\be \label {3.5a}
V(\B x_1, \B x_2) = \frac{g_s^2}{4 \pi} \, \frac{1}{|\B x_1 - \B x_2|} = \frac{g_s^2}{4 \pi} \, \frac{1}{x_{12}},
\ee
which is the non-relativistic limit of the one-gluon exchange interaction (in coordinate representation).
It reflects the $O(g_s^2)$ ``Coulombic" contribution to the interquark  potential. 
Note that it depends  only on the distance, $x_{12}$, between the points $\B x_1$ and $\B x_2$, as expected.
\vskip .3cm
Similarly, the component of the action corresponding to $\cL_{I_3}$, eq. \re{1-12}, is, in  first order,
%
\be \label{3.6}
\cA_{I_3}^{(1)} = \int \D{4}{x} \, \cL^{(1)}_{I_3} (x) =
-g_s \, f_{a\,b\,c} \int \D{4}{x} \, A^{(a)}_{(1)\,\nu}(x)
\, A_{(1)}^{{(b)}\,\mu} (x) \, \di_\mu A_{(1)}^{{(c)}\,\nu} (x).
\ee
This term corresponds to an $O(g_s^4)$ contribution to the energy. The
Green function $D$ appears in degree 3 in this expression,
which reflects the 3-gluon vertex interaction corresponding to the ``degree 3 in $A^{(a)}_\mu$" form of
$\cL^{(1)}_{I_3}$.
\par
In the static case, $A^{(a)}_{(1)\,\mu} (x) = \int \D{4}{x'} \, D(x-x')\, j^{(a)}_{\psi\,\mu} (x')$
becomes
%
\be \label{3.7}
A^{(a)}_{(1)\,\mu} (\B x) = \int \D{4}{x'} D(x-x') j^{(a)}_{\psi\,\mu} (\B x')
= \int \D{3}{x'} j^{(a)}_{\psi\,\mu} (\B x') \int \D{}{t'} D(x-x') =
\int \frac{\D{3}{x'}}{4\pi} \, \frac{j^{(a)}_{\psi\,\mu} (\B x')}{|\B x - \B x'|},
\ee
where we have used \re{3.4}, and thus
%
\be \label{3.8}
\di_t  A^{(a)}_{(1)\, \nu} (\B x) = 0, ~~~\di_k  A^{(a)}_{(1)\, \nu} (\B x) =
\int \frac{\D{3}{x'}}{4\pi}\,j^{(a)}_{\psi\,\nu} (\B x') \,\frac{\di}{\di x^k}
\frac{1}{|\B x - \B x'|}
= -\int \frac{\D{3}{x'}}{4\pi}\,j^{(a)}_{\psi\,\nu} (\B x')\,
\frac{x_k - x'_k}{|\B x - \B x'|^3},
\ee
where $k=1, 2, 3$.
Therefore, in the static case,
%
\be \label{3.9}
\cA_{I_3}^{(1)} = -\int dt \, H_{I_3}^{(1)}  =
-g_s \, f_{a\,b\,c} \int \D{4}{x} \,A^{(a)}_{(1)\,\nu}(\B x)
\, A_{(1)}^{{(b)}\,k} (\B x) \, \di_k A_{(1)}^{{(c)}\,\nu} (\B x).
\ee
Thus, by \re{3.7} and \re{3.8},
%
\bea
H_{I_3}^{(1)} &=& - \frac{g_s}{(4\pi)^3}\, f_{a\,b\,c}\int \D{3}{x}
\left[\int \D{3}{x_1}\frac{j^{(a)\,\nu}_\psi (\B x_1)}{|\B x - \B x_1|} \right]
\left[ \int \D{3}{x_2}\frac{j_{\psi\,k}^{(b)}(\B x_2)}{|\B x - \B x_2|} \right]
\left[ \int \D{3}{x_3}j^{(c)}_{\psi\,\nu} (\B x_3)
\frac{x^k - x_3^k}{|\B x - \B x_3|}\right]
\nn \\
&=& - \frac{g_s}{(4\pi)^3}\, f_{a\,b\,c}\int \D{3}{x_1}\,\D{3}{x_2}\,\D{3}{x_3} \,
j^{(a)}_{\psi\,\nu} (\B x_1) \,j_{\psi}^{(b)\,k} (\B x_2) \, j^{(c)\,\nu}_\psi (\B x_3)
\, U_k(\B x_1,\,\B x_2,\,\B x_3), \label{3.10}
\eea
where
%
\be \label{3.11}
U_k (\B x_1,\B x_2,\B x_3) = \int \D{3}{x} \, \frac{x^k - x^k_3}{|\B x - \B x_1||\B x - \B x_2||\B x - \B x_3|^3}
 =  \frac{\di}{\di x_3^k} \, U^{(3)}(\B x_1,\B x_2,\B x_3),
\ee
and
%
\be \label{3.112}
U^{(3)}(\B x_1,\B x_2,\B x_3) = -\int \, \frac{\D{3}{x}}{|\B x - \B x_1||\B x - \B x_2||\B x - \B x_3|}
 = -\int \, \frac{\D{3}{v}}{|\B v||\B v - \B x_{21}||\B v - \B x_{31}|};
\ee
here $\B x_{mn} = - \B x_{nm} \equiv \B x_m - \B x_n$.
Note that the integral in \re{3.112} is divergent and so must be regularised (this is discussed in ref. \cite{DD2}).
However the derivatives \re{3.11} are well-behaved (finite).
\par
We see that the interaction term \re{3.10} corresponds to a three-point potential,
%
\be \label{3.11a}
V_{k} (\B x_1, \B x_2, \B x_3) = - \frac{g_s^4}{(4 \pi)^3} \, U_k (\B x_1, \B x_2, \B x_3)
 = - \frac{g_s^4}{(4 \pi)^3} \, \frac{\di}{\di x_3^k} U^{(3)}(\B x_1, \B x_2, \B x_3), ~~k =1,2,3,
\ee
which is an $O(g_s^4)$ first-iterative -order ``correction" to the two-point Coulombic interaction $V(\B x_1, \B x_2)$
given in eq. \re{3.5a}.
\par
Lastly, the component of the action corresponding to $\cL_{I_4}$, eq. \re{1-13}, is, in  first-iterative-order
%
\bea \label{3.12}
\cA_{I_4}^{(1)} = \int \D{4}{x} \, \cL_{I_4} (x) =
-\frac{1}{4} g_s^2\,  f_{a\,b\,c} \,  f_{a\,d\,e} \int \D{4}{x} \, A^{(b)}_{(1)\,\mu} (x)
\, A^{(c)}_{(1)\,\nu} (x) \, A_{(1)}^{{(d)}\, \mu}(x) \, A_{(1)}^{{(e)}\,\nu}(x),
\eea
which is of $O(g_s^6)$ and of degree 4 in $D$, corresponding to a four gluon interaction vertex.
In the static case, this term becomes
%
\be \label{3.13}
\cA_{I_4}^{(1)} = -\frac{1}{4} g_s^2\,  f_{a\,b\,c} \,  f_{a\,d\,e} \int \D{4}{x} \,
A^{(b)}_{(1)\,\mu} (\bx)
\, A^{(c)}_{(1)\,\nu} (\bx) \, A_{(1)}^{{(d)}\, \mu}(\bx) \, A_{(1)}^{{(e)}\,\nu}(\bx)
= - \int \D{}t \, H^{(1)}_ {I_4} ,
\ee
where, by \re{3.7},
%
\bea \label{3.14}
H^{(1)}_ {I_4} &=& \frac{1}{4} g_s^2\,  f_{a\,b\,c} \,  f_{a\,d\,e} \int \D{3}{x} \,
A^{(b)}_{(1)\,\mu} (\B x) \, A^{(c)}_{(1)\,\nu} (\B x)
\, A_{(1)}^{{(d)}\, \mu}(\B x) \, A_{(1)}^{{(e)}\,\nu}(\B x)
\\ \nn
&=& \frac{g_s^2}{4(4\pi)^4} f_{a\,b\,c} \,  f_{a\,d\,e}
\inta \D{3}{x_1}\D{3}{x_2}\D{3}{x_3}\D{3}{x_4}
j^{(b)}_{\psi\,\mu} (\B x_1) \, j^{(c)}_{\psi \,\nu} (\B x_2)
\, j_\psi^{{(d)}\, \mu}(\B x_3) \, j_\psi^{{(e)}\,\nu}(\B x_4)
U^{(4)}(\B x_1, \B x_2, \B x_3, \B x_4),
\eea
with
%
\be \label{3.15}
\hspace{-0.8ex}
U^{(4)}(\B x_1, \B x_2, \B x_3, \B x_4) = \inta  \frac{\D{3}{x}}
{|\B x-\B x_1||\B x-\B x_2||\B x-\B x_3||\B x-\B x_4|}
= \inta  \frac{\D{3}{v}}{|\B v||\B v-\B x_{21}||\B v-\B x_{31}||\B v-\B x_{41}|}.
\ee
If we include the coupling constants (and related factors)
the corresponding potential-energy function is
%
\be \label{3.15a}
V_{I_4} (\B x_1, \B x_2, \B x_3, \B x_4) =
\frac{1}{4}\,\frac{g_s^6}{(4\pi)^4} \, U^{(4)}(\B x_1, \B x_2, \B x_3, \B x_4).
\ee
This is an $O(g_s^6)$ first-order-iterative four-point potential correction to the Coulombic two-point potential
\re{3.5a}.
\par
Unfortunately, the integrals \re{3.11}, \re{3.112} and \re{3.15} that define the three and four point potentials
cannot, in general, be evaluated explicitly, that is they cannot be expressed in terms of common analytic functions
(at least we do not know how to do so). Nevertheless
various general properties of these first-iterative-order non-Abelian corrections to the Coulombic inter-quark
potential can be readily established, and analytical expressions can be obtained for particular situations.
\par
The general properties and representations of $U_k$ and $U^{(4)}$ are presented and discussed in the Appendix,
where it is shows, inter alia, that
\par
1).~
The ``potentials" $U^{(3)}(\B x_1, \B x_2, \B x_3)$ and  $U^{(4)}(\B x_1, \B x_2, \B x_3, \B x_4)$
are, in fact, functions of the distances $x_{mn} = |\B x_m - \B x_n|$ only (as might be expected of a closed system).
\par
2).~
The evaluation of $U_k(\B x_1, \B x_2, \B x_3)$ can be reduced to the computation of two single quadratures and that
of $U^{(4)}(\B x_1, \B x_2, \B x_3, \B x_4)$ to a double quadrature, which must be done numerically.

\par
As mentioned, the three and four point functions $U^{(3)}$ and $U^{(4)}$ can be evaluated analytically for some particular cases.
Thus, when all three distances are equal, \ie $x_{12} = x_{13} = x_{23} = r$, the (regularised) three-point function
of eq. \re{3.11a}
becomes $U^{(3)}(r) =    4 \pi\ln (r/a)$, where $a$ is an arbitrary distance scale \cite{DD1,DD2}.
Similarly, when points $\B x_2$ and $\B x_3$ are coincident, \ie $x_{23}=0$ and $x_{12} = x_{13}$,
$~U^{(3)}(x_{12}) =    4 \pi \ln (x_{12}/a)$, and so
$$\ds U_k(\B x_1, \B x_2, \B x_3=\B x_2) \equiv
\left.\partial U^{(3)}(\B x_1, \B x_2, \B x_3)/\partial x_3^k\right|_{\sB x_3=\sB x_2}=
\ha\partial U^{(3)}(\B x_1, \B x_2, \B x_2)/\partial x_2^k=
- 2 \pi \frac{(\B x_1 - \B x_2)_k}{|\B x_1 - \B x_2|^2}.$$
This shows that the corresponding correction to the Coulombic one-gluon exchange potential \re{3.5a} due to the ``cubic"
interaction term \re{3.11}, in the non-relativistic limit, is of the form (cf. \re{3.11a})
%
\be \label{3.17}
 V_{k}(\B x_1, \B x_2, \B x_3=\B x_2) = \frac{g_s^4}{2(4 \pi)^2} \, \frac{(\B x_1 - \B x_2)_k}{|\B x_1 - \B x_2|^2}.
\ee
\par
Similarly, for the particular case $\B x_1 = \B x_3$ and $\B x_2 = \B x_4$, there is only one distance, $x_{12}$,
between the two pairs of coincident points, and the four-point potential function $U^{(4)}$, eq. \re{3.15},
and so $V_{I_4}$, eq. \re{3.15a}, can be evaluated explicitly:
%
\be \label{3.18}
U^{(4)}(x_{12}) = \int  \frac{\D{3}{v}}{|\B v|^2|\B v+\B x_{12}|^2} = \frac{\pi^3}{x_{12}},
~~~~{\rm hence} ~~~ V_{I_4}(x_{12}) = \frac{g_s^6}{4^5 \pi}\, \frac{1}{x_{12}}.
\ee
Although the expressions \re{3.17} and \re{3.18} are only segments of the three and four point
potentials $V_{k}(\B x_1, \B x_2, \B x_3)$, eq'n \re{3.11a},  and $V_{I_4}(\B x_1,\B x_2,\B x_3,\B x_4)$,
eq'n \re{3.15a}, they suggest that their behaviour is Coulomb-like in general. Note that these corrections are
$O(g_s^4)$ and $O(g_s^6)$ respectively.
%
%
\renewcommand{\theequation}{4-\arabic{equation}}
\setcounter{equation}{0}
\section{Concluding remarks}
We have used an approximate, iterative solution of the non-linear classical equations of motion of QCD
to derive expressions for the interaction terms corresponding to the
non-Abelian terms \re{1-12} and \re{1-13} of the QCD action.
In first iterative order, cf. equation \re{2-4}, these turn out to be expressions involving products of three and four
one-gluon exchange Green functions, corresponding to three- and four-gluon interaction vertices (cf. eq'ns
\re{3.6} and \re{3.12} respectively).
\par
We have examined these non-Abelian terms in the static, non-relativistic limit and found them to be
three- and four-point static potentials, \re{3.11a} and \re{3.15a}, that depend on the inter point coordinates only.
Although we could not express these three- and four-point potentials in terms of common analytic functions in general,
we could do so for some restricted sections and these indicate that the potentials $V_k(\B x_1,\B x_2,\B x_3)$
and $V^{(4)}(\B x_1,\B x_2,\B x_3,\B x_4)$ are Coulomb-like in general.

The derived three- and four-point cluster corrections together with the one-gluon exchange
interaction could be used as a short-range contribution in potential models of
baryons, tetra-quarks etc.
\par
The quantized theory, based on the reduced Lagrangian \re{2-6}, can be used to derive relativistic
few-quark equations, as was done for the scalar theory with non-linear mediating fields \cite{DD1, DD2}.
This shall be left for future work.


\section*{Appendix. Properties and evaluation of the three- and four-point potentials}
\renewcommand{\theequation}{A-\arabic{equation}}
\setcounter{equation}{0}

Components of the three-point potential \re{3.11} form the vector
potential:
%
\begin{equation}\lab{A.1}
\B U(\B x_1,\B x_2,\B x_3)
= \frac{\partial}{\partial\B x_3}U^{(3)}(\B x_1,\B x_2,\B x_3),
\end{equation}
where the function $U^{(3)}(\B x_1,\B x_2,\B x_3)$ (see eq.\re{3.112})
is studied in [4]. Particularly useful
is the following representation this function:
%
\begin{eqnarray}\lab{A.3}
U^{(3)}(\B x_1, \B x_2, \B x_3)&=&-\frac1{\pi^{3/2}}
\inta\D3k\inta\D3x\,{\rm e}^{-k_1^2(\sB x-\sB x_1)^2-k_2^2(\sB x-\sB x_2)^2-k_3^2(\sB x-\sB x_3)^2}\nn\\
&=&-\inta\frac{\D3k}{k^3}\,{\rm e}^{-(k_1^2k_2^2x_{12}^2+k_2^2k_3^2x_{23}^2+k_1^2k_3^2x_{13}^2)/k^2}\nn\\
&=&-\inta\D{2}{\hat{k}}\int\limits^{\infty}_0\frac{\D{}k}{k}\,
{\rm e}^{-(\hat k_1^2\hat k_2^2x_{12}^2+\hat k_2^2\hat k_3^2x_{23}^2+\hat k_1^2\hat k_3^2x_{13}^2)k^2},
\end{eqnarray}
where $x_{mn}\equiv|\B x_m-\B x_n|$ ($m,n=1,2,3$), $\hat{k}=\B k/k$, $k=|\B k|\equiv\sqrt{k^2_1+k^2_2+k^2_3}$
and $\int\D{2}{\hat k}$ denotes   
integration over the unit sphere in 3D $k$-space.
The integral \re{A.3} (as well as \re{3.112}) is divergent and needs to be regularized.
One way is to split it
into two terms: $U^{(3)}=\tilde U+U_0$,              
where $\tilde U(x_{12}, x_{23}, x_{13})$ is a finite function
of three scalar arguments and $U_0 = U(a, b, c)$ is an ``infinite constant" ($a, b, c$ are arbitrary constants) [4].

    Inserting \re{A.3} into r.h.s. of \re{A.1} discards the infinite constant $U_0$ and yields
the formula:
%
\begin{eqnarray}\lab{A.4}
\B U&=&\frac{\partial U^{(3)}}{\partial\B x_3}=
-\frac{\partial}{\partial\B x_3}\int\D{2}{\hat{k}}\int\limits^{\infty}_0\frac{\D{}k}{k}\,
{\rm e}^{-(\hat k_1^2\hat k_2^2x_{12}^2+\hat k_2^2\hat k_3^2x_{23}^2+\hat k_3^2\hat k_1^2x_{31}^2)k^2}\nn\\
&=&2\int\D{2}{\hat{k}}\hat k^2_3\left(\hat k^2_1\B x_{31}+\hat k^2_2\B x_{32}\right)
\int\limits^{\infty}_0\D{}k\,k\,{\rm e}^{-(\hat k_1^2\hat k_2^2x_{12}^2+\cdots)k^2}
\equiv\B x_{31}I_1+\B x_{32}I_2,
\end{eqnarray}
where
%
\begin{equation}\lab{A.5}
I_n=
\int\frac{\D{2}{\hat{k}}\,\hat k_n^2\hat k_3^2}
{\hat k_1^2\hat k_2^2x_{12}^2+\hat k_2^2\hat k_3^2x_{23}^2+\hat k_1^2\hat k_3^2x_{13}^2}, \quad
n=1,2.
\end{equation}

    Next, we introduce angular variables $\{\vartheta,\varphi\}$ on the unit sphere in 3D $k$-space, so that
$\hat k_1=\sin\vartheta\cos\varphi$, $\hat k_2=\sin\vartheta\sin\varphi$, $\hat k_3=\cos\vartheta$. Then
%
\begin{equation}\lab{A.6}
I_1=\int\limits_0^{2\pi}\D{}{\varphi}\cos^2\!\varphi\,J, \qquad
I_2=\int\limits_0^{2\pi}\D{}{\varphi}\sin^2\!\varphi\,J,
\end{equation}
where
%
\begin{eqnarray}
J&=&\int\limits_0^{\pi}\frac{\D{}{\vartheta}\sin\vartheta\cos^2\vartheta}{
(x_{12}\sin\vartheta\cos\varphi\sin\varphi)^2+\cos^2\!\vartheta(x_{13}^2\cos^2\!\varphi+x_{23}^2\sin^2\!\varphi)}\nn\\
&=&\frac{8}{x_{12}^2 R^2}\left[1-\frac{\sin2\varphi}{R}\mathrm{arctan}\,\frac{R}{\sin2\varphi}\right]
\lab{A.7}\\
\mbox{and}&&
R=\sqrt{(\cos2\varphi+\xi)^2+\eta^2},\nn\\
&&\xi=\frac{x_{13}^2-x_{23}^2}{x_{12}^2},\quad
\eta^2=\frac{[(x_{13}+x_{23})^2-x_{12}^2][x_{12}^2-(x_{13}-x_{23})^2]}{x_{12}^4}.
\lab{A.8}
\end{eqnarray}
Inserting \re{A.7} into \re{A.6} and using the integration variable $s=\cos2\varphi$ yields the quadratures:
%
\begin{eqnarray}
I_1&=&\frac{8}{x_{12}^2}\int\limits_{-1}^{1}\frac{\D{}{s}}{R^2}\left[
\sqrt{\frac{1+s}{1-s}}-\frac{1+s}{R}\mathrm{arctan}\,\frac{R}{\sqrt{1-s^2}}\right],\nn\\
I_2&=&\frac{8}{x_{12}^2}\int\limits_{-1}^{1}\frac{\D{}{s}}{R^2}\left[
\sqrt{\frac{1-s}{1+s}}-\frac{1-s}{R}\mathrm{arctan}\,\frac{R}{\sqrt{1-s^2}}\right].
\lab{A.9}
\end{eqnarray}
Note that for the special case $x_{23}=0$, the expressions \re{A.4} to \re{A.9} yield the
result \re{3.17}.
In the general case, that is, for arbitrary values of $x_{12}$, $x_{13}$ and $x_{23}$,
the integrals \re{A.9} need to be evaluated numerically.

\begin{figure}[h]
\begin{center}
\includegraphics[scale=0.6,angle=270]{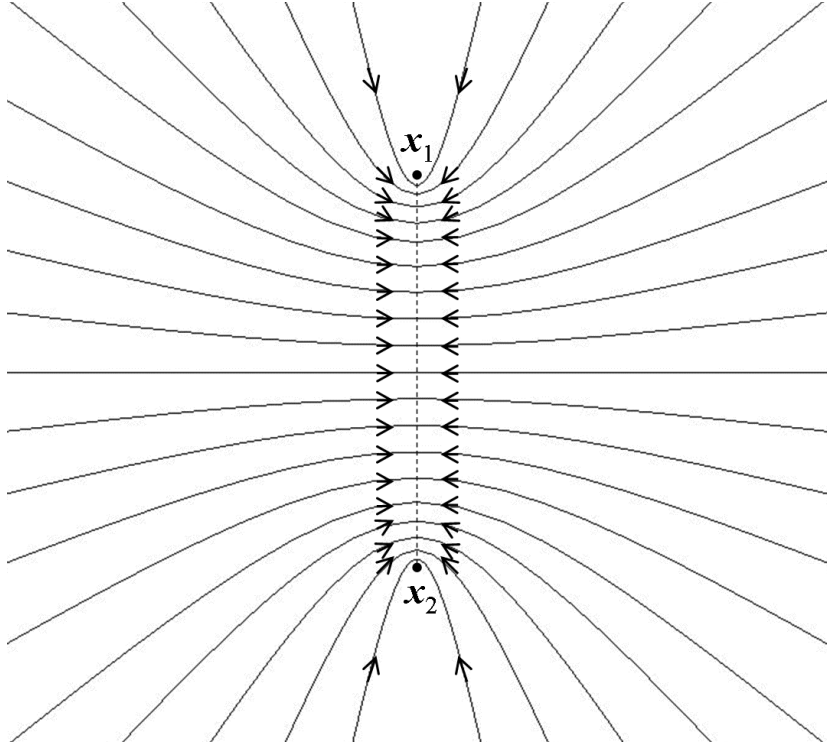}
\caption{Two-dimensional section
of the vector potential $\B U(\B x_1, \B x_2, \B x_3)$ as a
function of $\B x_3=\B r$ for fixed $\B x_1$ and $\B x_2$.
The arrows indicate the direction
of the vector field $\B U(\B x_1, \B x_2; \B r)$.
Note that it is invariant with respect rotations about the axis $\B x_1$--$\B x_2$.
The dashed line between $\B x_1$ and $\B x_2$ corresponds to $\B U=0$.}
\end{center}
\end{figure}

    The behavior of the three-point vector potential as a function of $\B x_3$
is illustrated in figure 1. We note the following symmetry properties of the vector potential:
\begin{itemize}
\item
translational invariance: $\B U(\B x_1+\Blambda, \B x_2+\Blambda,
\B x_3+\Blambda) = \B U(\B x_1, \B x_2, \B x_3)$, where
$\Blambda\in\Bbb R^3$;
\item
rotational covariance: $\B U({\rm R}\B x_1, {\rm R}\B x_2, {\rm R}\B x_3) =
{\rm R}\B U(\B x_1, \B x_2, \B x_3)$, where ${\rm R}\in{\rm SO(3)}$;
\item
partial permutational invariance: $\B U(\B x_2, \B x_1, \B x_3 )= \B U(\B x_1, \B x_2, \B x_3)$;
\item
scaling transformation: $\B U(\lambda\B x_1, \lambda\B x_2, \lambda\B x_3) =
\lambda^{-1}\B U(\B x_1, \B x_2, \B x_3)$, where $\lambda\in\Bbb R_+$.
\end{itemize}
These properties follow from the properties of the three-point scalar potential \re{3.112}
stated in \cite{DD1,DD2}. We also note  that the scaling transformation of the potential $\B U$
reflects its Coulomb-like behaviour.
\bigskip

The four-point scalar potential \re{3.15} can be treated in similar manner:
%
\begin{eqnarray}\lab{A.10}
U^{(4)}(\B x_1,\dots,\B x_4) &=& \!\int\!\frac{\D3x} {|\B x-\B x_1|\cdots|\B x-\B x_4|}=\frac1{\pi^2}
\inta\D4k\inta\D3x\,{\rm e}^{-k_1^2(\sB x-\sB x_1)^2-\cdots-k_4^2(\sB x-\sB x_4)^2}\nn\\
&=&\frac1{\sqrt{\pi}}\inta\D{3}{\hat{k}}\int\limits^{\infty}_0\D{}k\,
{\rm e}^{-X^2k^2}=\int\frac{\D{3}{\hat{k}}}{\sqrt{X^2}},\\
\mbox{where~~~}X^2&\equiv&
\hat k_1^2\hat k_2^2x_{12}^2+\hat k_1^2\hat k_3^2x_{13}^2+\hat k_1^2\hat k_4^2x_{14}^2
+\hat k_2^2\hat k_3^2x_{23}^2+\hat k_2^2\hat k_4^2x_{24}^2+\hat k_3^2\hat k_4^2x_{34}^2, \nn
\end{eqnarray}
$\hat k_n=k_n/k$ ($n=1,...,4$), $k=\sqrt{k^2_1+\cdots+k^2_4}$,
and $\int\D{3}{\hat k}$ denotes an integration over a unit hyper-sphere in 4D $k$-space.

    Next, we introduce angular variables $\{\chi,\vartheta,\varphi\}$ on the unit hyper-sphere in 4D $k$-space, so that
$\hat k_1=\sin\chi\sin\vartheta\cos\varphi$, $\hat k_2=\sin\chi\sin\vartheta\sin\varphi$,
$\hat k_3=\sin\chi\cos\vartheta$, $\hat k_4=\cos\chi$, and
$\int\D3{\hat k}=\int\limits_0^{2\pi}\D{}\varphi\int\limits_0^{\pi}\sin\vartheta\,\D{}\vartheta
\int\limits_0^{\pi}\sin^2\,\chi\,\D{}\chi$. Then using the integration variables $u=\cos\chi$, $v=\cos\vartheta$,
$w=\cos2\varphi$ reduces the integral \re{A.10} to the  form:
%
\begin{eqnarray}\lab{A.11}
U^{(4)}(x_{12},\dots,x_{34})&=&4\int\limits_{-1}^{1}\frac{\D{}w}{\sqrt{1-w^2}}
\int\limits_{0}^{1}\D{}v\, I,\\
\mbox{where~~~}I&=&
\int\limits_{0}^{1}\frac{\D{}u}{\sqrt{A^2+u^2B^2}}=\frac1B\ln\left(\frac{\sqrt{A^2+B^2}+B}{A}\right), \nn\\
A^2&=&\left\{\qu x^2_{12}(1-w^2)(1-v^2)+\ha[x^2_{13}+x^2_{23}+(x^2_{13}-x^2_{23})w]v^2\right\}(1-v^2),\nn\\
B^2&=&\ha[x^2_{14}+x^2_{24}+(x^2_{14}-x^2_{24})w](1-v^2)+x^2_{34}v^2-A^2\nn
\end{eqnarray}
This double integral can be evaluated numerically. Evaluation of \re{A.11} for the case $\B x_1 = \B x_3$ and
$\B x_2 = \B x_4$ gives the same result as \re{3.18}.

Finally we note, that the four-point potential \re{A.11} possesses the same translational invariance and
scaling transformation properties as the three-point one does.
Besides, it is invariant under arbitrary rotation and permutation of its arguments:
\begin{itemize}
\item
rotational invariance: $U^{(4)}({\rm R}\B x_1, {\rm R}\B x_2, {\rm R}\B x_3, {\rm R}\B x_4) =
U^{(4)}(\B x_1, \B x_2, \B x_3, \B x_4)$, where ${\rm R}\in{\rm SO(3)}$;
\item
complete permutational invariance:\\ $U^{(4)}(\B x_2, \B x_1, \B x_3, \B x_4)=
U^{(4)}(\B x_1, \B x_3, \B x_2, \B x_4)=\dots=U^{(4)}(\B x_1, \B x_2, \B x_3, \B x_4)$.
\end{itemize}
%
%
%

%
\end{document}